\documentclass[12pt]{article}
\usepackage{epsfig,rotating}
\usepackage{cite}

\newcommand{\beq}{\begin{equation}}
\newcommand{\eeq}{\end{equation}  }

\newcommand{\la}{\langle}
\newcommand{\ra}{\rangle}

\newcommand{\bec}{\begin{center}}
\newcommand{\eec}{\end{center}}

\def\d{\delta}

\def\E{\mbox{e}^+\mbox{e}^-}

\newcommand{\Wp}{\mbox{w}^+}
\newcommand{\Wn}{\mbox{w}^-}
\newcommand{\W}{\mbox{w}}
\newcommand{\WW}{\mbox{ww}}

\newcommand{\dl}{\Delta\rho (\pm,\pm)}
\newcommand{\du}{\Delta\rho (+,-)}

\newcommand{\WWqqqq}{\mathrm{W}^+\mathrm{W}^-\to
\bar{\mathrm{q}}_1\bar{\mathrm{q}}_2\mathrm{q}_3\mathrm{q}_4}

\newcommand{\WWqqln}{\mathrm{W}^+\mathrm{W}^-\to
\bar{\mathrm{q}}\mathrm{q}l\bar{\nu}_l}

\begin{document}
\clearpage
\pagestyle{empty}
\setcounter{footnote}{0}\setcounter{page}{0}%
\thispagestyle{empty}\pagestyle{plain}\pagenumbering{arabic}%

\hfill ANL-HEP-PR-98-84

\hfill HEN-411

\hfill June 1998 

\vspace{1.0cm}

\begin{center}

\vskip 0.8in plus 2in

{\Large\bf Bose-Einstein Correlations and  Color Reconnection \\ 
in W-pair production \\[-1cm]}

\vspace{2.0cm}

{\large S.~V.~Chekanov $^a$ 
\footnote[1]{
On leave from
the Institute of Physics,  AS of Belarus,
Skaryna av.70, Minsk 220072, Belarus.}, 
E.~A.~De ~Wolf $^b$ 
\footnote[2]{Onderzoeksdirecteur FWO (Vlaanderen), Belgium.}, 
W.~Kittel $^c$}

{\begin{itemize}
\itemsep=-1mm

\normalsize 
\item[$^a$]

\small
Argonne National Laboratory,
9700 S.Cass Avenue,
Argonne, IL 60439
USA

\normalsize
\item[$^b$]

\small
Department of Physics, Universitaire  Instelling Antwerp,
B-2610 Wilrijk, Belgium

\normalsize
\item[$^c$]
 
\small
High Energy Physics Institute Nijmegen
(HEFIN), University of Nijmegen/NIKHEF,
NL-6525 ED Nijmegen, The Netherlands
 
\end{itemize}
}

\normalsize
\vspace{2.0cm}

\begin{abstract}
We propose a systematic  study of Bose-Einstein correlations
between identical hadrons coming from different W decays. Experimentally 
accessible signatures of these correlations as well as of possible  color 
reconnection effects are discussed on the basis of two-particle inclusive 
densities. 
\end{abstract}

\end{center}
\newpage
\setcounter{page}{1}
\section{Introduction}

One of the important problems  in the study of the $\E$ annihilation
at LEP2 energies is the understanding of  production and decay of W-boson
pairs. Due to the fact that hadrons originating from  W decay overlap
in space and are created in time almost simultaneously, it is natural
to expect that there are correlations between hadrons originating  from
different W decays due to color reconnection  and Bose-Einstein (BE)
interference. These effects may affect the accuracy with which the W mass
could be measured \cite{sjo,kh,t1}.  

The DELPHI Collaboration has  estimated both effects \cite{d1,d2}. 
At present level of statistics, no evidence for these effects has
been found. However,  no systematic theoretical  
treatment of the BE effect 
in W-pair production has been given so far.

The problem of BE correlations cannot be separated from the color 
reconnection  effect. For the color reconnection phenomenon, theoretical 
model investigations have recently been performed 
\cite{t2,t3,t4,t4s}. For example, it was proposed 
in \cite{t2} to measure a difference between the mean hadron multiplicity 
in four-jet final states ($\WWqqqq$) and twice the hadronic  multiplicity 
in two-jet events ($\WWqqln$). Having clear advantages at the present level 
of low statistics, this method, however, cannot be sensitive to all possible
correlations  which may exist due to cross-talk between hadrons and
may be experimentally accessible in the near future. 

In this paper we present a systematic study of both  effects leading to a  
stochastic  dependence between hadrons coming from different W decays. Our 
study is mainly limited to a discussion of 
two-particle inclusive densities but can easily be generalized
to higher-order correlations. 

\section{Independent W-pair decay}

\subsection{Many-particle inclusive description}

In this subsection we shall give a very general formalism of independent
WW decay using  generating functionals for many-particle inclusive densities
(see \cite{kkk} for a review). 

A  distribution of final-state
particles produced in four-jet WW decay in a phase-space
domain $\Omega$ is fully determined by 
the generating functional 
\beq
\mathcal{R}^{\WW} [u(p)] = 1 + \sum_{n=1}^{\infty} \frac{1}{n!}
\int_{\Omega}\rho^{\WW} (p_1,p_2,\ldots , p_n) 
u(p_1)\ldots u(p_n)
\prod_{i=1}^{n} \mathrm{d}p_i , 
\label{x22}
\eeq
where $\rho^{\WW} (p_1,p_2,\ldots , p_n)$ is 
the $n$-particle inclusive distribution  with
$p_i$ being  the 4-momentum of $i$th particle.
The inclusive densities can be recovered  
from the functional differentiation of (\ref{x22})
\beq
\rho^{\WW} (p_1,p_2,\ldots , p_n) = \partial^n  \mathcal{R}^{\WW} [u(p)] /
\partial u(p_1)\ldots \partial u(p_n) \mid_{u=0}. 
\eeq
Since high-order inclusive densities contain redundant information
from  lower-order  densities, it is advantageous to consider
the $n$-particle (factorial)
cumulant correlation functions $C^{\WW} (p_1,p_2, \ldots , p_n)$ 
which are obtained from the generating functional
\beq
\mathcal{G}^{\WW} [u(p)] = \ln \mathcal{R}^{\WW} [u(p)],
\label{pp1}
\eeq
so that 
\beq
C^{\WW} (p_1,p_2,\ldots , p_n) = \partial^n  \mathcal{G}^{\WW} [u(p)] /
\partial u(p_1)\ldots \partial u(p_n) \mid_{u=0}.
\label{pp1a}
\eeq

Analogously, one can define the generating functionals for the final-state
hadrons in two-jet WW decay,
\beq
\mathcal{R}^{\W} [u(p)] = 1 + \sum_{n=1}^{\infty} \frac{1}{n!}
\int_{\Omega}\rho^{\W} 
(p_1,p_2,\ldots , p_n) 
u(p_1)\ldots u(p_n)
\prod_{i=1}^{n} \mathrm{d}p_i,
\eeq

\beq
\mathcal{G}^{\W} [u(p)] = \ln \mathcal{R}^{\W} [u(p)]
\label{pp2}
\eeq
with $\rho^{\W}(p_1,p_2,\ldots , p_n)$ being the $n$-particle inclusive
density for  two-jet WW decay.

Let us consider an uncorrelated WW decay  scenario. In this we assume
that each W boson showers and fragments into final-state hadrons without
any reference to what is happening to the other.
In this case $\mathcal{R}^{\WW} [u(p)]$ is the  
product of the  generating functionals for the two-jet WW decay
of differently charged W's
\beq
\mathcal{R}^{\WW} [u(p)] =\mathcal{R}^{\Wp} [u(p)] 
\> \mathcal{R}^{\Wn} [u(p)]. 
\label{den}
\eeq
In terms of the generating functionals for the correlation functions,
this can be represented as follows
\beq
\mathcal{G}^{\WW} [u(p)] =\mathcal{G}^{\Wp} [u(p)] + 
\mathcal{G}^{\Wn} [u(p)]. 
\label{com}
\eeq
We  explore the relation (\ref{den}) only for the two-particle
inclusive density. Being a simple characteristic
beyond single-particle inclusive spectra, it  is  this  quantity 
that  is very often used in the correlation analysis, 
especially in connection with the BE interference.

\subsection{Two-particle inclusive density}

Let us first define the two-particle inclusive density $\rho (1,2)$ 
for particles $1$ and $2$ in the 
variable $Q_{12}=\sqrt{-(p_1-p_2)^2}$ as
\beq
\rho (1,2)=\frac{1}{N_{\mathrm{ev}}}
\frac{\mathrm{d}n_{\mathrm{pairs}}}{\mathrm{d}Q_{12}}, \qquad
\int_{Q}\rho (1,2)\mathrm{d}Q_{12}=\la n_1(n_2-\delta_{12})\ra . 
\label{q2}
\eeq 
Here, $N_{\mathrm{ev}}$ is the number of events (in a theoretical
limit $N_{\mathrm{ev}}\to\infty$), 
$n_{\mathrm{pairs}}$ is the number of particle pairs,    
$n_1$ is the number of particles of type $1$ in the event,
$n_2$ that of type $2$. For different hadrons 
(or identical hadrons coming from different events) 
$\d_{12}=0$ and  $\d_{12}=1$ 
for identical  hadrons coming from the same event.
Since there are 
two possible  combinations for identical hadrons, positive-positive and 
negative-negative, we combine  both samples  into a single  one with a  
factor $1/2$. Hereafter, we shall refer  to this  as the like-charged 
particle sample  and will symbolize it as $(\pm,\pm)$. For  unlike-charged 
particle combinations, we adopt the notation $(+,-)$. The 
integration is performed in (\ref{q2}) over the full range of $Q$  of the
variable $Q_{12}$, so that
\beq
\la n_1(n_2-1)\ra\equiv F_2,
\label{q3}
\eeq
where $F_2$ is the second-order (unnormalized) factorial moment for full
phase space.

The single-particle and two-particle inclusive densities
in $Q_{12}$ variable for the four-jet  WW hadronic decay can directly 
be obtained performing two successive functional differentiations  of 
(\ref{den}) over the probing function $u(p)$,
\beq
\rho^{\WW}(1)=\rho^{\Wp} (1) + \rho^{\Wn} (1), 
\label{r1x}
\eeq
\beq
\rho^{\WW}(1,2)=\rho^{\Wp}(1,2) + \rho^{\Wn}(1,2)
+2\rho^{\Wp}(1)\rho^{\Wn}(2).  
\label{k11} 
\eeq
Note that the latter expression 
differs  from the sum of two-particle 
densities for each independent source taken separately. 

Performing the same  functional differentiations of (\ref{com}),
one can find the two-particle correlation
function in the  four-jet WW decay,
\beq
C^{\WW}(1,2)=C^{\Wp }(1,2) + C^{\Wn }(1,2), 
\label{r1}
\eeq 
This illustrates the fact that, in contrast to the two-particle densities, 
the correlation functions are additive and  do not contain 
the contribution from  lower-order inclusive densities. 

Experimentally,  it is advantageous to  rewrite (\ref{k11}): 
\beq
\rho^{\WW}(1,2)=\rho^{\Wp}(1,2) + \rho^{\Wn}(1,2)
+2\rho^{\Wp\Wn}_{\mathrm{mix}}(1,2),  
\label{k1} 
\eeq
where we replaced $\rho^{\Wp}(1)\rho^{\Wn}(2)$ with  
the track mixing two-particle density 
$\rho^{\Wp\Wn}_{\mathrm{mix}}(1,2)$ obtained 
by pairing particles from  
different two-jet WW  events, to insure that particles coming
from differently charged W's do not correlate. This  technique  leads
to factorization of $\rho^{\Wp\Wn}_{\mathrm{mix}}(1,2)$ into
the product of the single-particle densities. 

Let us consider different charged-particle combinations. 
Following (\ref{k1}), one can define  
\beq
\dl \equiv \rho^{\WW}(\pm,\pm) - 2\,\rho^{\W}(\pm,\pm)
- 2\rho^{\Wp\Wn}_{\mathrm{mix}}(\pm,\pm),
\label{6t}
\eeq
\beq
\du \equiv \rho^{\WW}(+,-)-2\, \rho^{\W}(+,-)
-2\rho^{\Wp\Wn}_{\mathrm{mix}}(+,-),
\label{7t}
\eeq
where we assume that
\beq
\rho^{\W}(+,-)\equiv\rho^{\Wp}(+,-)=\rho^{\Wn}(+,-),
\eeq
\beq
\rho^{\W}(\pm,\pm)\equiv\rho^{\Wp}(\pm,\pm)=\rho^{\Wn}(\pm,\pm).
\eeq
Expressions (\ref{6t}) and (\ref{7t}) are evidently equal to zero for
uncorrelated  four-jet WW decay.

One can integrate (\ref{r1x}) and 
(\ref{k11}) over the $Q$ interval to obtain the  relations for average
multiplicity and second-order 
factorial moment in  uncorrelated four-jet WW decays:
\beq
\Delta \equiv \la n_{\WW} \ra -\la n_{\Wp} \ra - \la n_{\Wn} \ra =0,
\label{aww}
\eeq
\beq
\Delta F_2 \equiv F_2^{\WW} - F_2^{\Wp} - F_2^{\Wn} 
- 2\la n_{\Wp}\ra \la n_{\Wn}\ra =0 .
\label{ff1}
\eeq
The latter  equation can also  be directly   
obtained from the assumption on uncorrelated 
WW decay. Indeed, taking into account (\ref{q3}), equation (\ref{ff1}) can 
be rewritten as 
\beq
\la n_{\WW}^2 \ra - \la n_{\WW}\ra -\la n_{\Wp}^2 \ra + 
\la n_{\Wp}\ra - \la n_{\Wn}^2 \ra + 
\la n_{\Wn}\ra - 2\la n_{\Wp}\ra \la n_{\Wn}\ra =0.
\label{ff2}
\eeq
Assuming that for each WW event $n_{\WW}=n_{\Wp}+n_{\Wn}$ and, for 
uncorrelated WW decay, 
$\la n_{\Wp} n_{\Wn} \ra =\la n_{\Wp}\ra \la n_{\Wn}\ra$, 
one can see that the left-hand side of this equation  is indeed zero. Note 
that in this particular case all $\la n_{\W}^2\ra$ terms cancel, i.e.,  
equation (\ref{ff2}) holds for any full-phase-space multiplicity distribution. 
 
By construction, 
\beq
\Delta F_2=\int_{Q} \Delta \rho (1,2),
\eeq
omitting the charge dependence for simplicity. 

A deviation of $\dl$,  $\du$  or $\Delta F_2$ from zero 
is possible only in the case
of correlated WW decay. It is very important to note, however, that the 
opposite is not true: $\dl = \du =0$ is a 
necessary, but not a {\em sufficient} condition for uncorrelated WW 
decay.
This is further illustrated in the 
appendix using generating functions. 

\subsection{Reduced BE correlations in independent WW decay}

A commonly acceptable method
to study the BE effect is based on the calculation
of the following correlation function:
\beq
R(1,2)=\frac{\rho (1,2)}{\rho(1)\rho (2)}=
1 + \frac{C(1,2)}{\rho(1)\rho (2)}.
\label{b1}
\eeq
In this it is assumed that the two-particle density $\rho (1,2)$ for 
identical (like-charged) boson combinations contains no additional 
correlations except those connected with the BE interference. 

Experimentally, the reference sample $\rho(1)\rho (2)$ is usually 
constructed by using the track mixing method of pairing identical particles 
from different events. To make it possible to estimate the BE effect in the 
case when some extra correlations are present in $\rho (1,2)$, $R$ should 
be further divided by the same function but calculated from Monte Carlo (MC) 
models without the BE effect. This technique  is based on the assumption
that different types of correlations can be  factorized and Monte Carlo models
are able  to describe all other possible correlations correctly.

Another way to estimate $R$ is to use the reference sample composed of 
unlike-charged particles from the same event.  This  method is affected by
the presence of dynamical correlations due to the decay products of resonances.

A first attempt to describe the BE correlations in four-jet WW decay would 
be to understand the behavior of $C(1,2)$ when there is no stochastic 
dependence between W pairs. This can be done if one remembers that the overall
topology of $\WWqqln$ events is quite similar to that of Z boson decay at LEP1
energies. Therefore, one can assume that the correlation function 
$C^{\mathrm{W}}(1,2)$ in the two-jet WW events  is the same as the 
correlation  function $C^{\mathrm{Z}}(1,2)$ in Z boson decay. From 
(\ref{b1}) one can write
\beq
\rho^{\W}(1,2)=\rho^{\W}(1)\rho^{\W}(2)+C^{\mathrm{Z}}(1,2) .
\eeq
Substituting this into (\ref{k11})  assuming 
$\rho^{\WW}(1)=2\,\rho^{\W}(1)$ {\em  for full overlap in $Q_{12}$}, 
one has  
\beq
\rho^{\WW}(1,2)=\rho^{\WW}(1)\rho^{\WW}(2)+
2\,C^{\mathrm{Z}}(1,2)
\eeq
and
\beq
R^{\WW}(1,2) = 1 + \frac{1}{2} 
\frac{C^{\mathrm{Z}}(1,2)}{\rho^{\W}(1) 
\rho^{\W}(2)} .
\eeq
From this follows the  fact that, in the absence of WW correlations,
the strength of the BE correlations in four-jet WW events is only
half of the strength in Z boson  or in two-jet W decay!
In practice, the overlap will not be complete, even in the particular
projection variable used, and the suppression will be less severe
in actual Monte Carlo simulation below, but the point is that not the
same, but a reduced BE
effect has to be expected for WW events even in the absence of
inter-W correlations.
Note that the  possibility of a decrease of the BE effect in the case of 
independent four-jet WW decay  has already been pointed out in \cite{t4},
but  without quantitative  estimates  of this effect.

Of course, the latter  conclusion is  correct   only in the case of no
correlations between hadrons  coming from different W bosons. We shall discuss
the degree of validity of this assumption in the next subsection.   

\subsection{Monte Carlo study}

To check the validity of $\dl=\du=0$ in  
(\ref{6t}) and (\ref{7t}),  we use the PYTHIA 6.1 
Monte Carlo  model \cite{jet} with the L3  default parameters \cite{jetd} 
for LUND hadronization without  BE correlations.\footnote{We use the L3 
default  since in this paper we study the model with the two sets of 
parameters -  with and without the BE simulation. Both models  have been 
tuned  to reproduce the same global-shape variables and single-particle 
densities at Z peak energy.} A cut on charged-particle multiplicity 
$N_{\mathrm{ch}}> 2$ is used. The total number of events is 4000 for four-jet 
and 8000 for two-jet WW decays generated  at c.m. energy of 190 GeV. 
Since the hadronic multiplicity of two-jet events is 
affected by $\tau$ decays, hadrons from $\tau$ decays are excluded.
For the given statistics and tuning, the average charged-particle 
multiplicity is $\la n_{\W}\ra=16.90\pm 0.05$ for two-jet ($\WWqqln$) and 
$\la n_{\WW}\ra =33.6\pm 0.1$ for four-jet $\WWqqqq$ decay. This is smaller 
than for the original JETSET default  since long-lived resonances (such as  
K$^0$, $\Lambda$) are declared to be stable. From the mean  multiplicities, 
one obtains the ratio $\la n_{\WW}\ra /2\la n_{\W}\ra =0.994\pm 0.004$.

To obtain $\rho^{\Wp\Wn}_{\mathrm{mix}}$ in the track-mixing method, 
we generate the particle  multiplicity 
$N_{\mathrm{p}}$ according to the Poisson 
distribution with the mean obtained from two-jet WW events.
Then, we generate events using  $N_{\mathrm{p}}$ tracks from two-jet 
WW events  imposing the constraint  that each track should originate from 
a different event. We require the total charge of
the generated event to be zero, and that the two particles of the pair
originate from differently charged W's. 
In addition, for a given generated event 
with the multiplicity $N_{\mathrm{p}}$, only tracks from an original event
of multiplicity $N_{\mathrm{p}}-4\le N \le N_{\mathrm{p}}+4$ are 
used. The analysis is based on 250k track-mixed events. 
 
\begin{figure}[htbp]
\vspace{-1.5cm}
\begin{center}
\mbox{\epsfig{file=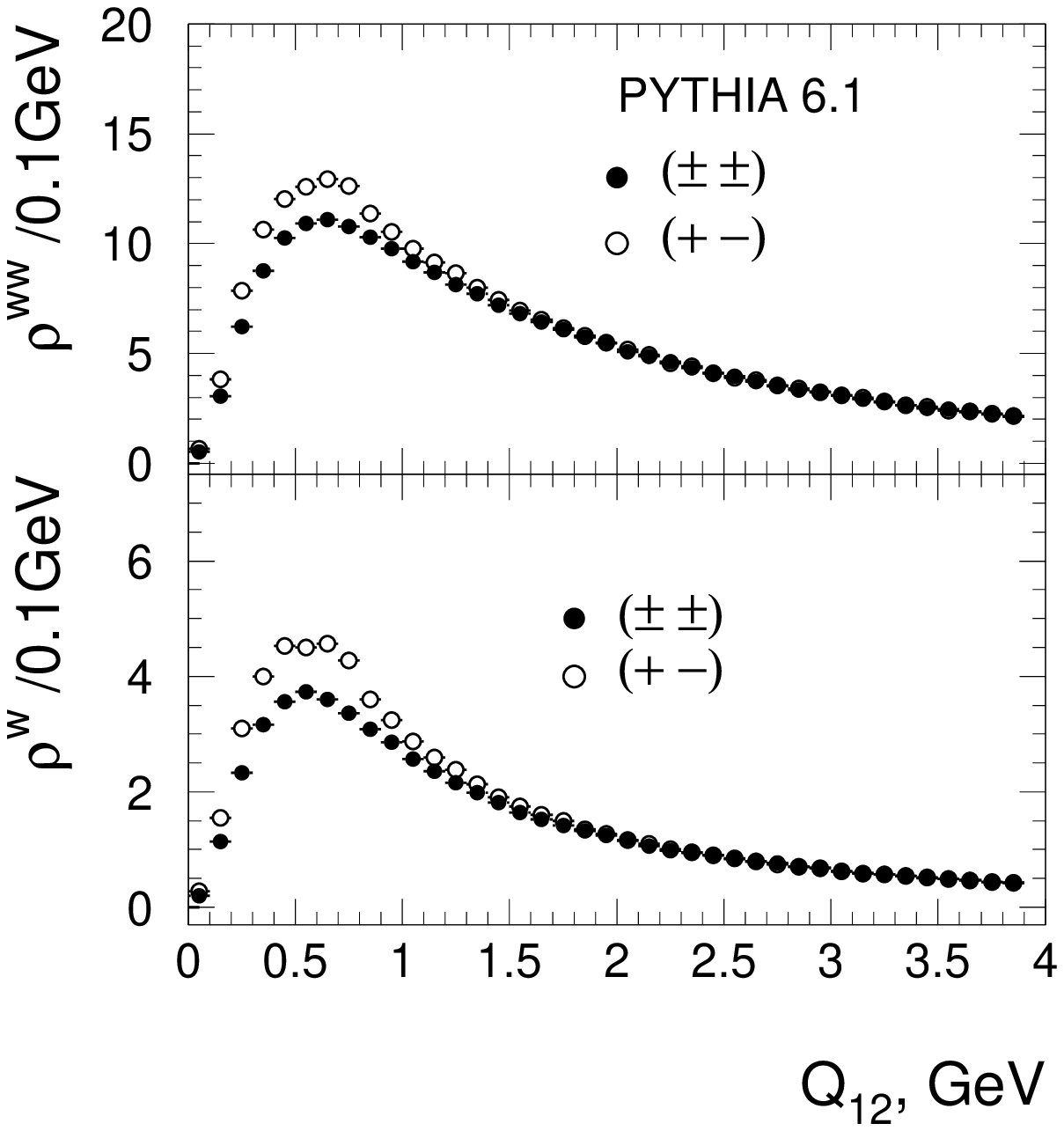, width=0.9\linewidth}}
\end{center}
\caption{
Two-particle inclusive densities for four-jet and two-jet WW decays 
generated with PYTHIA  MC without BE correlations.}
\label{1j}
\end{figure}

\begin{figure}[htbp]
\vspace{-2.0cm}
\begin{center}
\mbox{\epsfig{file=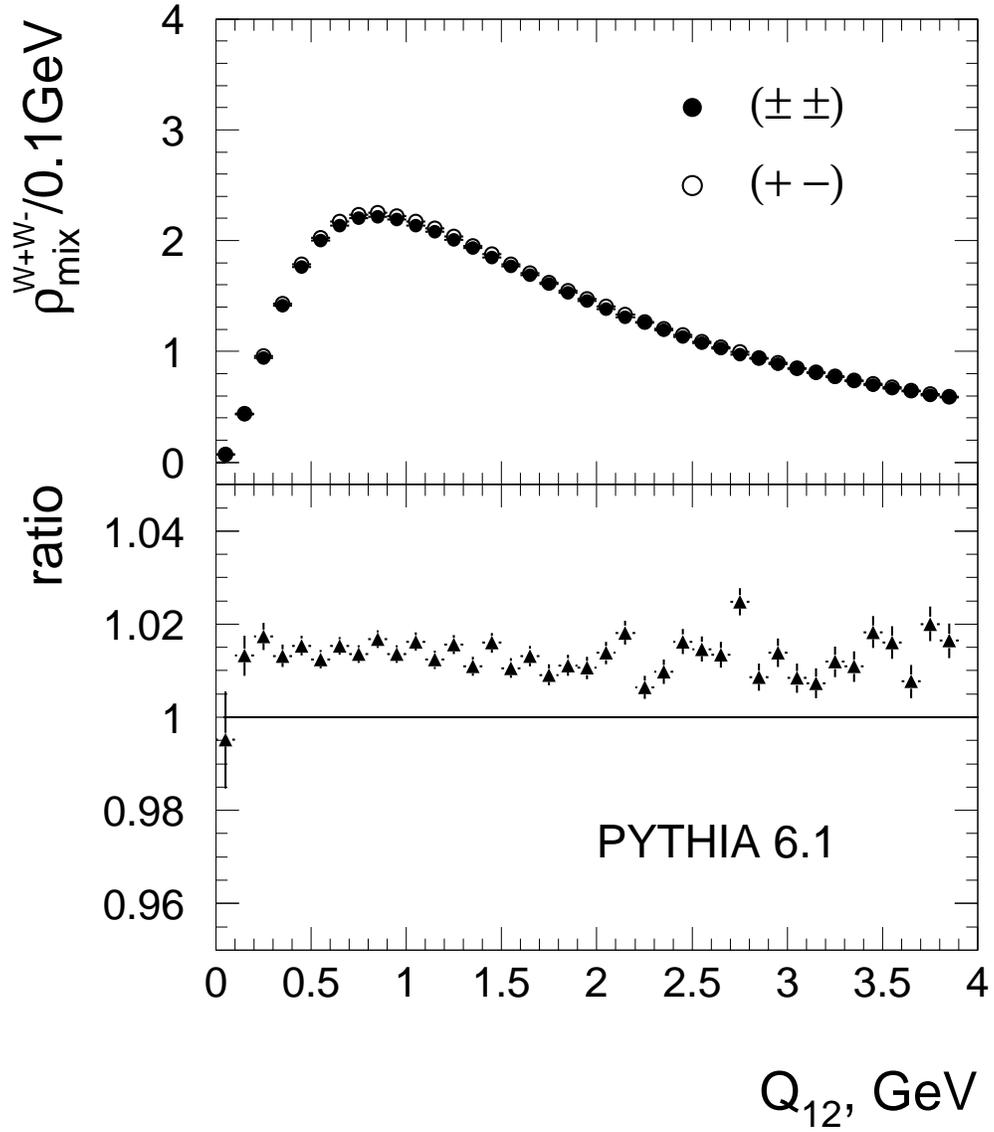, width=0.9\linewidth}}
\end{center}
\caption{
Two-particle densities obtained with the track mixing method.
Since it is difficult to  distinguish between different charged combinations,
the figure also shows the ratio
$\rho^{\Wp\Wn}_{\mathrm{mix}}(\pm ,\pm)/\rho^{\Wp\Wn}_{\mathrm{mix}}(+,-).$} 
\label{2j}
\end{figure}

\begin{figure}[htbp]
\vspace{-1.5cm}
\begin{center}
\mbox{\epsfig{file=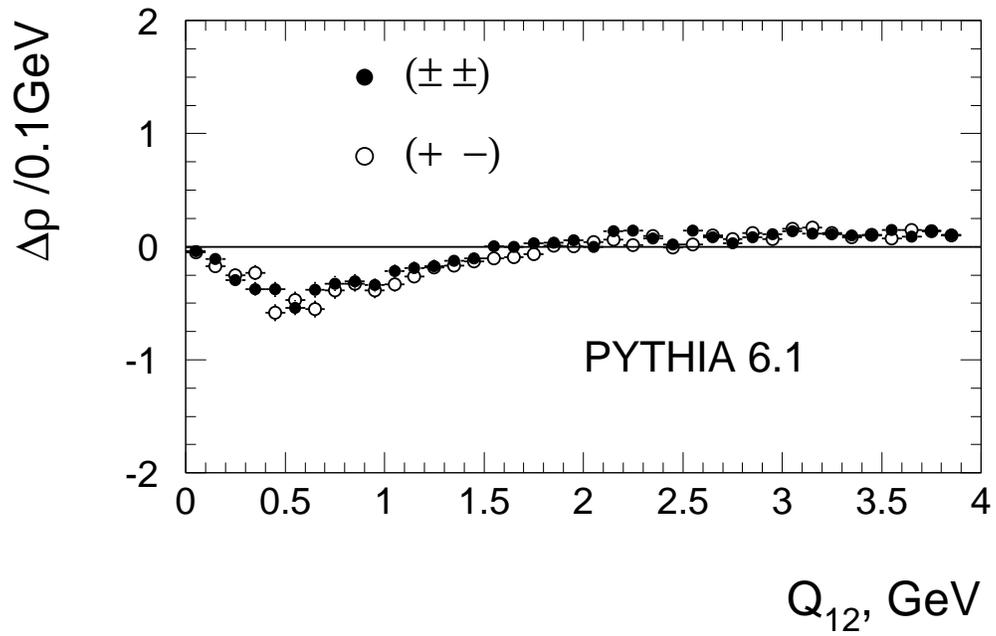, width=0.9\linewidth}}
\end{center}
\caption{
$\Delta\rho$ obtained with
PYTHIA  MC without BE correlations.}
\label{3j}
\end{figure}

Figs.~\ref{1j} and ~\ref{2j} show the behavior of the three terms 
in (\ref{6t}) and (\ref{7t}). Since the results  
for $\rho^{\Wp\Wn}$$(\pm,\pm)$ and $\rho^{\Wp\Wn}_{\mathrm{mix}}$$(+,-)$ 
are nearly on top of each other, Fig.~\ref{2j} also  shows  the ratio 
$\rho^{\Wp\Wn}_{\mathrm{mix}}(\pm,\pm)/\rho^{\Wp\Wn}_{\mathrm{mix}}(+,-)$.
Finally, Fig.~\ref{3j} shows $\dl$ and $\du$. 
As seen the assumption of 
independent hadronic W decay does not hold, even when  color reconnection 
and BE effects are not included in the MC code. The degree of such 
non-independence can be estimated by integrating the left hand-side of 
(\ref{6t}) and (\ref{7t}) over full phase space to obtain $\Delta F_2$ 
(\ref{ff1}): 
\beq
\int_Q\Delta\rho(\pm,\pm)\simeq 
\int_Q\Delta\rho(+,-)\simeq -4.3 \pm 0.3.
\label{integ1}
\eeq

Note that to simplify our study, we evaluated statistical errors assuming
that there are no correlations between points at different $Q_{12}$ values. 
This is a strong assumption, especially for the $Q_{12}\sim 0.5$ region
where the contribution of resonances is largest and phase-space points  
are strongly correlated. In addition, we did not take into account
that the average multiplicity $\la n_{\W}\ra$ has its error, which should 
be taken into account when  generating the particle multiplicity 
$N_{\mathrm{p}}$ for $\rho^{\Wp\Wn}_{\mathrm{mix}}$ according to the Poisson 
distribution. Therefore, the  values of errors shown in the figures are 
lower limits.

Statistical errors for $\rho^{\Wp\Wn}_{\mathrm{mix}}$  are rather 
small. Indeed, statistics available 
for the calculations of $\rho^{\W}$ is 
determined by the number of pairs. This is proportional to
$\la n\ra^2_{\W}N_{\mbox{ev}}$, where $N_{\mbox{ev}}$ is the number
of two-jet events. However, if we do not take into account the cut 
$N_{\mathrm{p}}-4\le N \le N_{\mathrm{p}}+4$, 
$\rho^{\Wp\Wn}_{\mathrm{mix}}$ is roughly determined by
$\la n\ra^2_{\W}N_{\mbox{ev}}^2$ 
since tracks are taken from different events.

\begin{figure}[htbp]
\vspace{-1.5cm}
\begin{center}
\epsfig{file=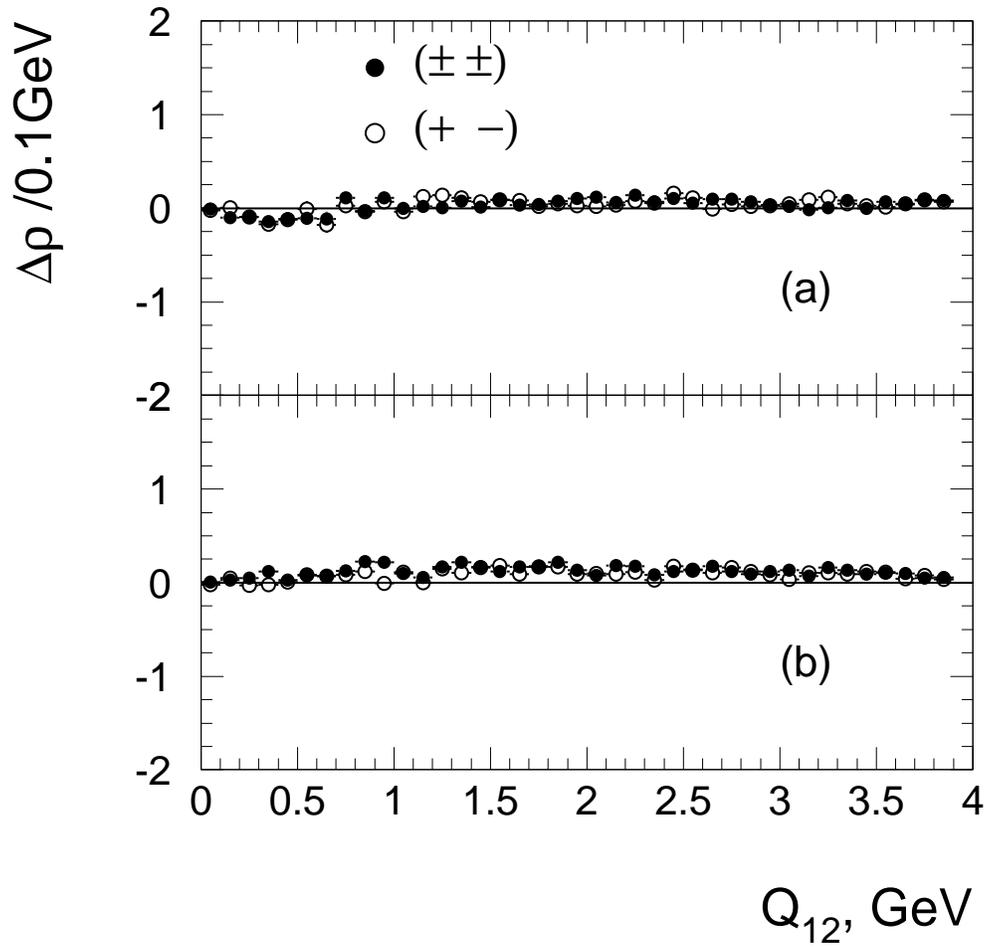, width=0.9\linewidth} 
\end{center}
\caption{
$\Delta\rho$ obtained with PYTHIA MC without BE correlations, 
combining a) two different
Z boson events into one single four-jet event; b) two two-jet events
of opposite W charge into a single four-jet event. \label{4a}\label{4b}}
\end{figure}

To check the correctness of the method, we simulated pseudo-W events
using   hadron production at the Z peak. The average multiplicity of 
these events  is rather similar to that of  two-jet WW decay. We combined  
two independent Z boson events generated with PYTHIA 6.1. Then, we 
considered  this hypothetical event to be a ``four-jet WW event''.
A single Z boson decay is considered to be a ``two-jet WW event''. Since Z 
boson decay products are taken from different events, they {\em a-priori} 
do not  correlate and (\ref{6t})  and (\ref{7t}) have to be  zero. 
Using the same 
program as that for the original WW sample, we repeated the previous 
calculations. The results are given in Fig.~\ref{4a}a. 

To see whether there is any effect from the fact that the two W's
in a four-jet event carry opposite charge, we repeated the above analysis
combining two two-jet events with different W charge 
into a single ``four-jet event''. 
The result is  given in  Fig.~\ref{4b}b.

In both figures,  Fig.~\ref{4a}a and ~\ref{4b}b,  
there is a small systematic
deviation from the zero line.
This can be due to residual
correlations which are not completely removed in the track-mixing
sample.
However, taking into account that
statistical errors are underestimated, such a deviation  is
rather small and will be neglected.

\section{Correlated WW production}

\subsection{General features}

From the MC study in the previous section it follows that the standard 
assumption of independent WW decay  is a rather naive simplification 
when we are dealing with the two-particle inclusive densities. One can 
consider a few possible reasons leading to non-independent WW decay:

\medskip

1) Energy-momentum conservation. Consider the production of two W's in 
the c.m.  frame. The mass of each W boson is distributed  according to the 
Breit-Wigner shape, i.e. for each event the two  masses are unequal and 
differ from the nominal W mass (which is 80.25 GeV for the L3 default in 
PYTHIA). From this it can be seen that there is a competition between 
the Breit-Wigner mass distribution and overall momentum rescaling to 
conserve the total energy $E_{\mathrm{cm}}$ and to allow for enough phase
space. 

\medskip

2) Apart from the Breit-Wigner distribution, the overall topology of WW
events is generated according to the matrix element approach with the 
nominal W mass. This calculation includes Coulomb interaction between 
different W's \cite{jet}.
Theoretical calculations of the Coulomb effects on the  WW production
can be found in \cite{col}. 

\medskip

3) Since spin information is included into the matrix elements,
there are angular correlations. 

\medskip

While the contribution from the two last effects is not well  understood 
yet and, presumably, is small, the first effect is most important since it 
may produce negative correlations: The overall shape of the multiplicity 
distribution in the four-jet WW decay is slightly narrower than that 
expected  from naive  superposition of two independent two-jet WW decays.

From the MC study, one can estimate the degree of
(linear) stochastic dependence between two W masses.
For this one can calculate the correlation coefficient,
\beq
r(m_+,m_-)=\left[\sigma^2(m_+)\sigma^2(m_-)\right]^{-1/2}
\left( \la m_+m_-\ra - \la m_+\ra \la m_-\ra\right),
\eeq
\beq 
-1\le r(m_+,m_-)\le 1,
\eeq
where $\sigma^2(m_{\pm})$ is the variance and $\la\ldots\ra$ stands for
the average over all events. If there is no correlation between the two 
W masses $m_+$ and $m_-$, then $r(m_+,m_-)=0$.  
Our MC estimate gives  $r(m_+,m_-)\simeq -0.04\pm 0.02$, i.e., there is a 
small  negative correlation between masses. Since the multiplicity of 
particles is determined by $m_+$ and $m_-$, this means that a large hadron 
multiplicity from one W boson slightly  suppresses the multiplicity of 
hadrons coming from the other W.  

The effects discussed above  are  not the only phenomena  which can lead 
to non-zero values of (\ref{6t}) and (\ref{7t}). At hadronization scale 
distances, the space separation between  the two W decay vertices is 
rather small ($\sim 0.1$ fm) and the hadronization regions of the two 
W bosons overlap. For this system, soft partons originating from different 
W bosons are close-by in space and could form color-singlet clusters from 
which the observable final-state hadrons emerge \cite{web}. Therefore, the 
origin of these hadrons is difficult to determine. Such an effect, usually 
called color reconnection, could lead to an additional non-independency 
of W decay products. In terms of the LUND model, the reconnection occurs 
when strings overlap like for a type I superconductor or when they  cross 
like for a type II superconductor \cite{t4}. 

After the transformation of partons into hadrons, the BE correlations 
can give an additional contribution to the overall correlations,
since the space-time  separation between hadrons is still smaller than  
the typical source radii ($\sim 0.5-1.0$  fm) of the BE correlations. 

In the case of  the interference effects, one can assume 
\beq
\dl =\Delta\rho^{\mbox{ec}}(\pm,\pm) + \Delta\rho^{\mbox{be}}(\pm,\pm) +
\Delta\rho^{\mbox{cr}}(\pm,\pm),
\label{cr2}
\eeq
\beq 
\du =\Delta\rho^{\mbox{ec}}(+,-)+\Delta\rho^{\mbox{cr}}(+,-),
\label{cr1}
\eeq
where $\Delta\rho^{\mbox{ec}}$ is the contribution from energy conservation 
and other non-interference effects, $\Delta\rho^{\mbox{be}}$ represents the 
BE correlations and $\Delta\rho^{\mbox{cr}}$-color-reconnection correlations.

One can directly investigate
the interference effects by calculating  the difference:
\beq
\d\rho = \dl - \du . 
\label{dif}
\eeq
Since the track mixing terms are very similar, one has
\beq
\d\rho \simeq \rho^{\WW}(\pm,\pm) - 2\,\rho^{\W}(\pm,\pm) -
\rho^{\WW}(+,-)+2\, \rho^{\W}(+,-),
\label{dif1}
\eeq
which no longer involves  the track mixing terms since they cancel.
Taking into account
the fact that  $\Delta\rho^{\mbox{ec}}(\pm,\pm)$ and 
$\Delta\rho^{\mbox{ec}}(+,-)$
are the same  (see MC studies above), from (\ref{cr2}) and (\ref{cr1})
one can see that $\d\rho$ resolves  only the interference terms 
\beq
\d\rho \simeq  \Delta\rho^{\mbox{be}}(\pm,\pm) + 
\Delta\rho^{\mbox{cr}}(\pm,\pm)
   - \Delta\rho^{\mbox{cr}}(+,-).
\label{dif2}
\eeq
If the color-reconnection effects are charge-independent,
$\d\rho$ is fully determined by the BE correlations, 
$\d\rho \simeq \Delta\rho^{\mbox{be}}(\pm,\pm)$.

\subsection{BE correlations}

The study  of BE interference in the form of
an enhancement of the two-particle correlation function 
by comparing fully hadronic and double semi-leptonic events
has been proposed in \cite{sjo}. Following this method,
the DELPHI Collaboration measured  the following 
correlation function \cite{d1}:
\beq
R^*=\frac{\rho^{\WW}(\pm,\pm) - 2\,\rho^{\W}(\pm,\pm)}
{\rho^{\WW }(+,-) - 2\,\rho^{\W}(+,-)}.
\label{del1}
\eeq
Because of (\ref{6t}) and (\ref{7t}),  this expression  is equal to
\beq
R^*=\frac{2\rho^{\Wp\Wn}_{\mathrm{mix}}(\pm,\pm) +\dl}
{2\rho^{\Wp\Wn}_{\mathrm{mix}}(+,-) + \du}.
\label{dex}
\eeq
Note that (\ref{dex}) has very little 
to do with the standard definition of
the BE correlation function (\ref{b1}).

This can  be seen if
one assumes that $\du =0$ and
$\rho^{\Wp\Wn}_{\mathrm{mix}}(\pm,\pm)\simeq
\rho^{\Wp\Wn}_{\mathrm{mix}}(+,-)$,
\beq
R^* \sim  1+ \frac{\dl}
{2\rho^{\Wp\Wn}_{\mathrm{mix}}(\pm,\pm)}.
\label{dex1}
\eeq
Formally, the structure of (\ref{dex1}) is similar to (\ref{b1}). 
However, since $\dl$ is different from  $C(1,2)$ for identical 
pions originating from different W bosons, $R^*$  is not the BE  correlation 
function. For example, one can see that $R^*$ is always peaked at 
$Q_{12}\to 0$  for any  slow change  in $\dl$, since 
$\rho^{\Wn\Wp}_{\mathrm{mix}}(\pm,\pm)$ is a decreasing function 
for $Q_{12}\to 0$. In fact, $\Delta\rho (\pm,\pm)$
is non-dynamically distorted by this division.
Such  a distortion by the single-particle density in
(\ref{del1})-(\ref{dex1}) is properly removed in the
definition (\ref{dif1}) to study the interference effects.

\subsection{Monte Carlo studies}

To see how the BE correlations  affect $\delta\rho$ and $R^*$, 
we use the PYTHIA 6.1 Monte Carlo with the BE effect included
for all identical pions. 
The BE correlations are simulated with the LUBOEI model.
After the model retuning,  the average charged-particle
multiplicity is $\la n_{\W}\ra=16.72\pm 0.05$ for two-jet and 
$\la n_{\WW}\ra =33.5\pm 0.1$ for four-jet WW  decay.
The ratio is 
$\la n_{\WW}\ra /2\la n_{\W}\ra =0.997\pm 0.004$.

\begin{figure}[htbp]
\vspace{-1.5cm}
\begin{center}
\mbox{\epsfig{file=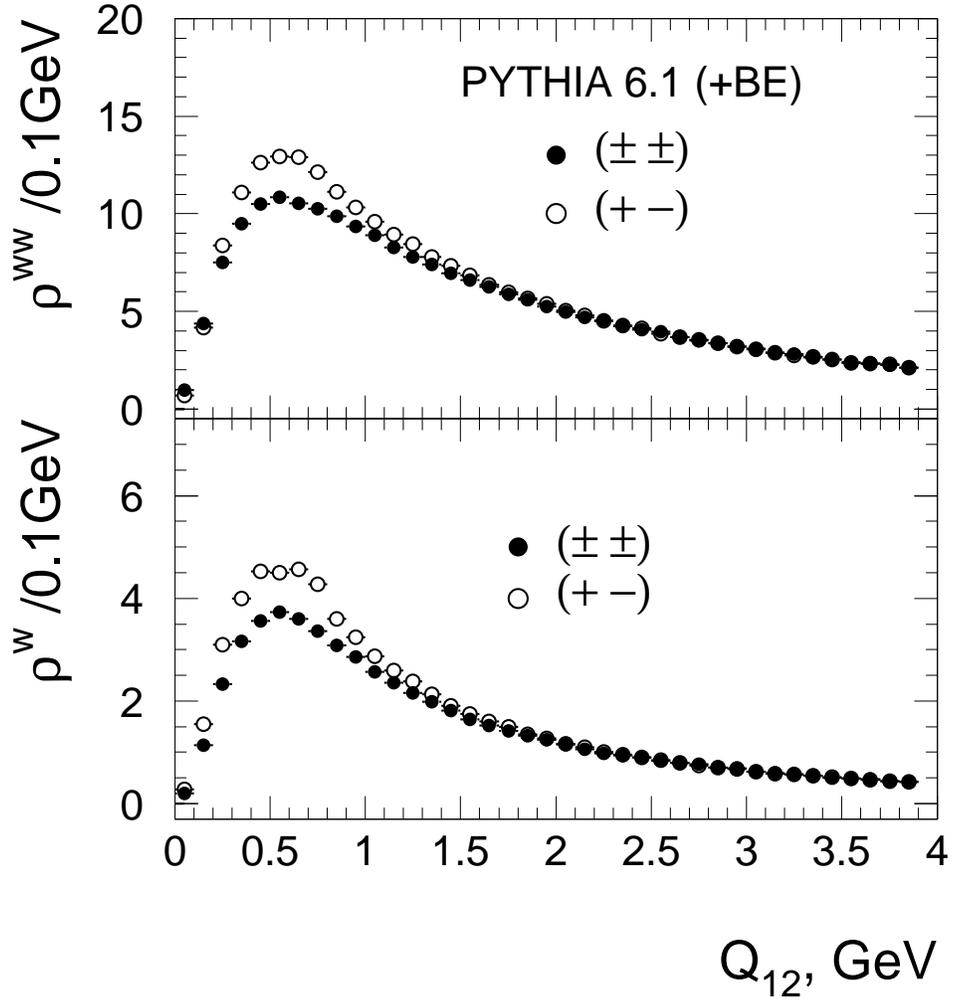, width=0.9\linewidth}}
\end{center}
\caption{
Two-particle inclusive densities for fully hadronic
and semi-leptonic WW decays generated with
PYTHIA MC with BE correlations included.}
\label{5j}
\end{figure}

\begin{figure}[htbp]
\vspace{-1.5cm}
\begin{center}
\mbox{\epsfig{file=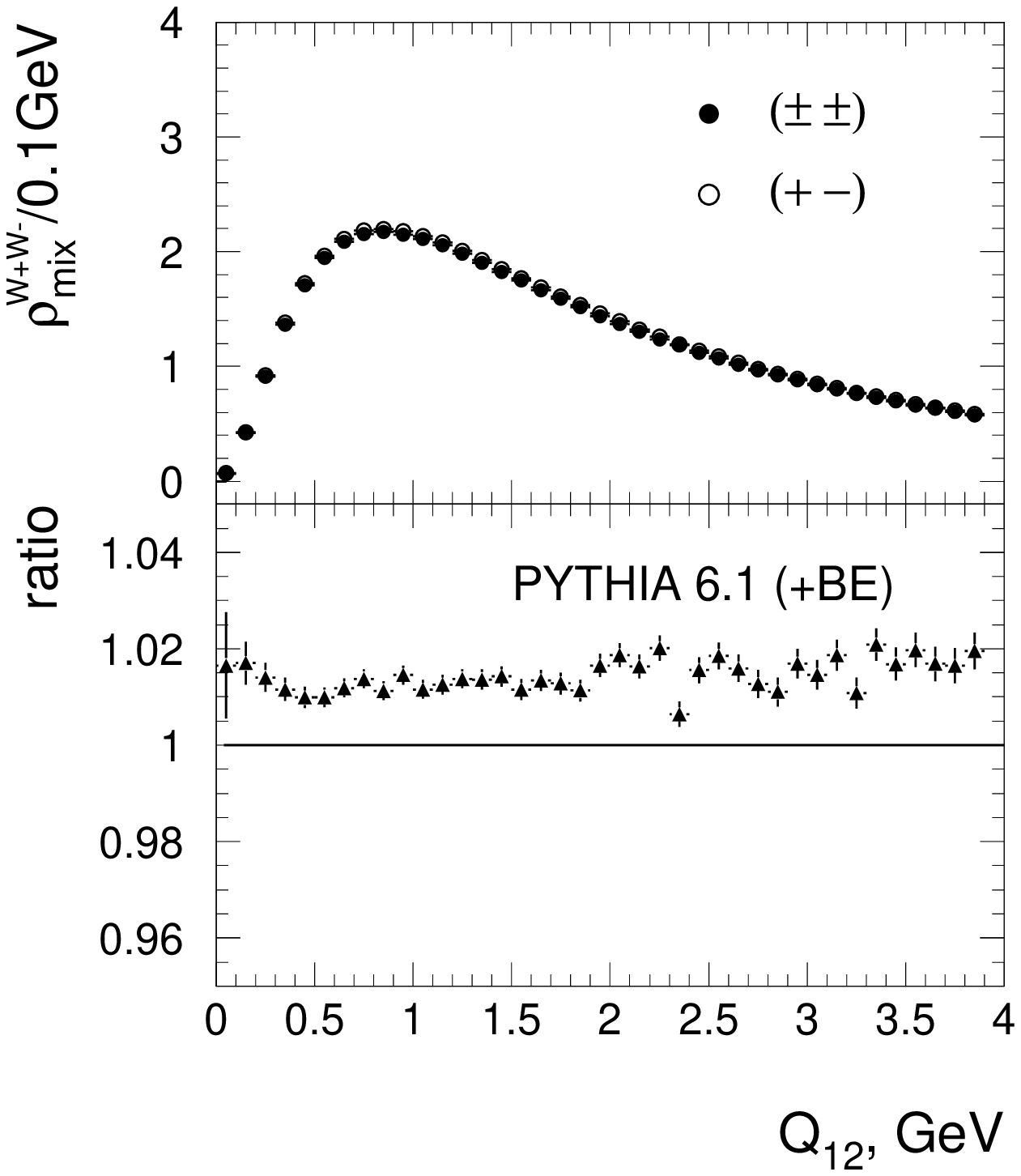, width=0.9\linewidth}}
\end{center}
\caption{
Two-particle densities obtained with the track mixing method for
PYTHIA  with BE correlations.
The figure also shows the ratio
$\rho^{\W\W}_{\mathrm{mix}}(\pm ,\pm)/\rho^{\W\W}_{\mathrm{mix}}(+,-).$}
\label{6j}
\end{figure}

\begin{figure}[htbp]
\vspace{-1.5cm}
\begin{center}
\mbox{\epsfig{file=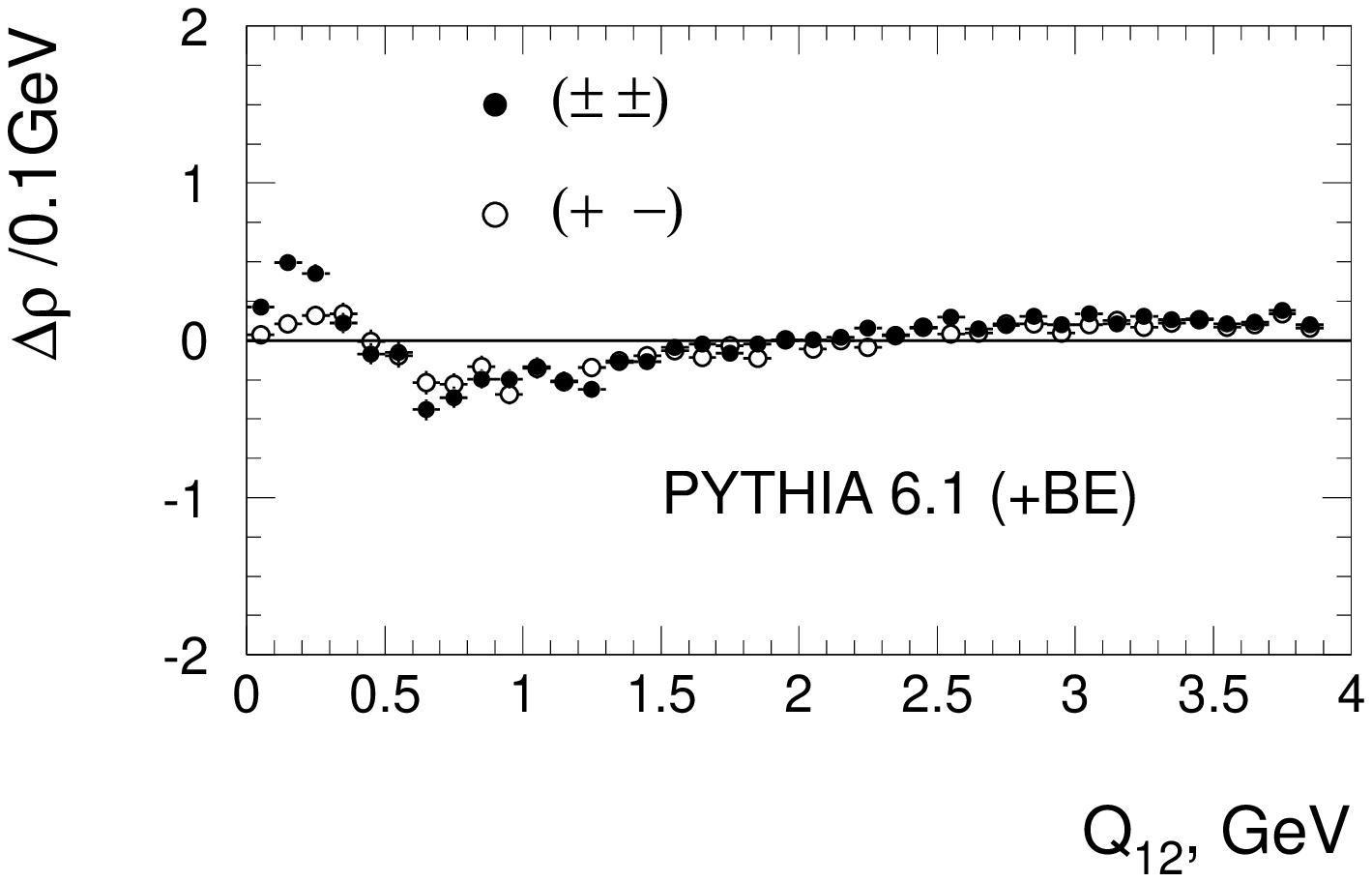, width=0.9\linewidth}}
\end{center}
\caption{
$\Delta\rho$ for PYTHIA  MC with BE correlations.}
\label{7j}
\end{figure}

Figs.~\ref{5j},~\ref{6j},~\ref{7j} show
the  terms  of (\ref{6t}) and (\ref{7t}) for  the case 
of the BE correlations, as they are implemented into the
Monte Carlo code. The most obvious difference is the
(expected) effect of the BE correlations on $\dl$ at small $Q_{12}$
in Fig.~\ref{7j} (c.f. Fig.~\ref{3j}). However,
also $\du$ is affected and non-zero in LUBOEI. The approximation
(\ref{dex1}), therefore, is not valid and (\ref{dex}) cannot measure the
standard BE correlation function.   
Integrating the left hand-side of
(\ref{6t}) and (\ref{7t}) over full phase space, 
one has (c.f. (\ref{integ1}))
\beq
\int_Q\Delta\rho(\pm,\pm)\simeq 1.54 \pm 0.04, \qquad
\int_Q\Delta\rho(+,-)\simeq 1.43 \pm 0.04
\label{integ2}.
\eeq

\begin{figure}[htbp]
\vspace{-1.5cm}
\begin{center}
\mbox{\epsfig{file=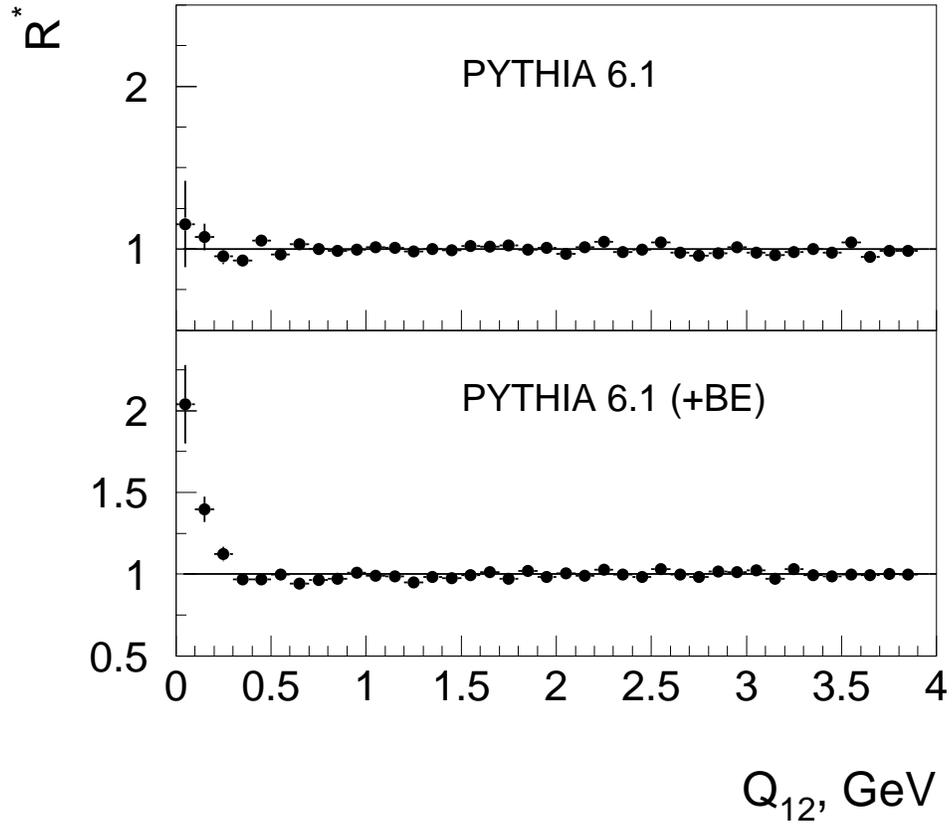, width=0.9\linewidth}}
\end{center}
\caption{
$R^*$ for PYTHIA  MC without and with  BE correlations.}
\label{8j}
\end{figure}

\begin{figure}[htbp]
\vspace{-1.5cm}
\begin{center}
\mbox{\epsfig{file=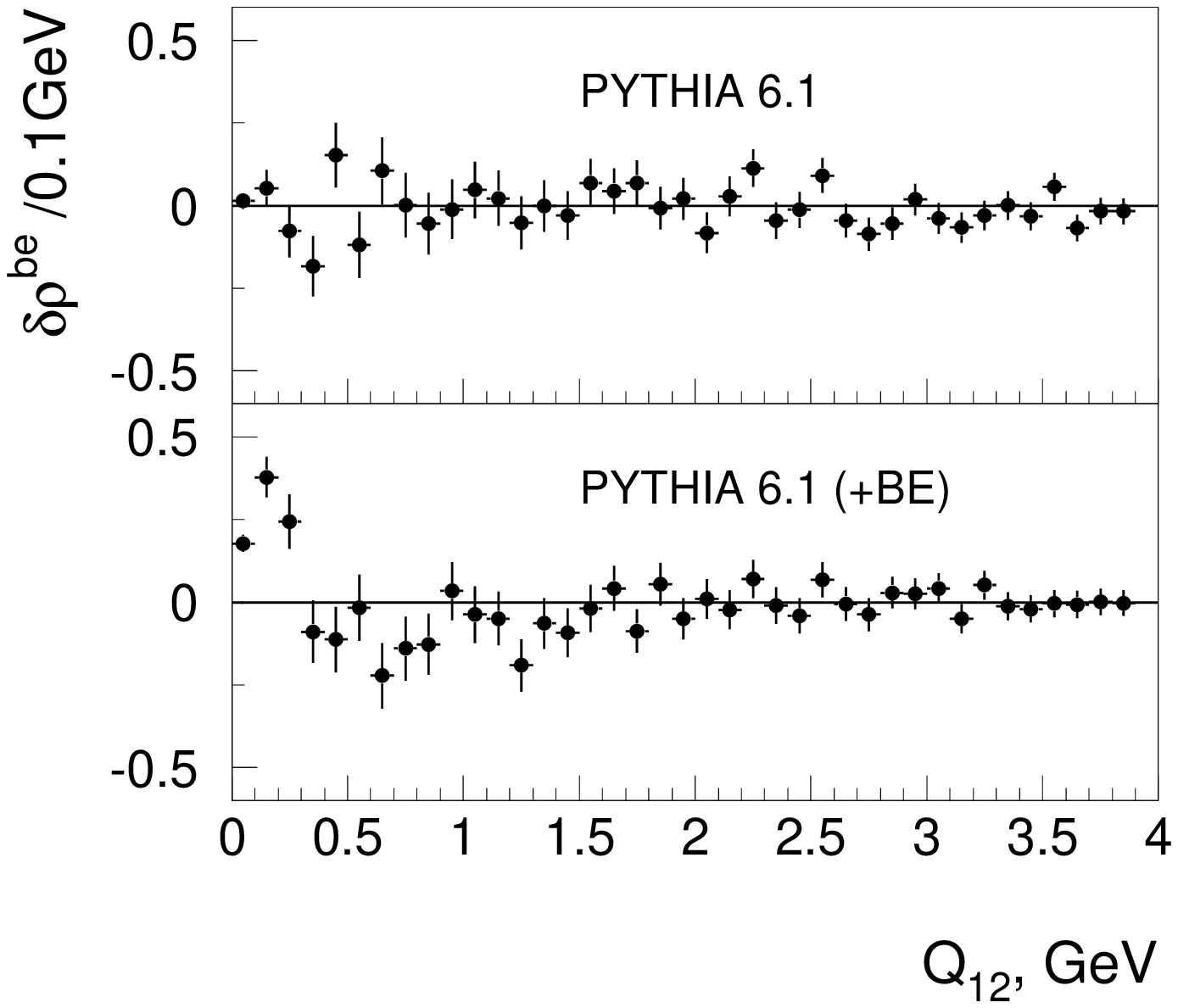, width=0.9\linewidth}}
\end{center}
\caption{
$\delta\rho$ for PYTHIA  MC without 
and with BE correlations.} 
\label{9j}
\end{figure}

Figs.~\ref{8j} and ~\ref{9j} show the behavior of $R^*$ and $\delta\rho$.
The BE effect appears stronger in $R^*$ than in  $\delta\rho$. In addition, 
statistical errors in Fig.~\ref{8j} are much smaller. However, as we have 
noted already, this is mainly because of the form  of 
$\rho^{\WW}_{\mathrm{mix}}(\pm,\pm)$ at small $Q_{12}$. This  leads to a  
behavior of $R^*$ appearing similar to that of BE correlations, even if 
$\dl$ is a small $Q_{12}$-independent constant. 

The inconsistency in the BE correlation study by means of $R^*$ can
be seen in Fig.~\ref{8j}. In the parameterization of the BE correlations, 
the L3 default is a spherical Gaussian source
$R(Q_{12})\sim 1 +\lambda\exp(-r^2Q^2_{12})$ with the
correlation strength parameter $\lambda=1.5$ and radius $r=0.6$ for
all 9 particle species. This means that the BE from 
different W bosons should have a similar form. However, Fig.~\ref{8j} 
shows that the shape is far from Gaussian.

The structure of the BE correlations between hadrons originating from
different W bosons can be observed from the study of $\delta\rho$, despite 
its evidently small signal. One can see from Fig.~\ref{7j} that LUBOEI 
changes the unlike-particle spectrum as well: Since LUBOEI spoils the  
overall energy-momentum conservation when it shifts identical particles 
to reproduce the expected two-particle correlation function,  momenta of 
non-identical particles are modified. 
Assuming that there is
no color reconnection, expressions (\ref{cr2}),  
(\ref{cr1}) and  (\ref{dif2}) are modified for LUBOEI as
\beq
\dl =\Delta\rho^{\mbox{ec}}(\pm,\pm) + 
\Delta\rho^{\mbox{be}}_{\mathrm{LUB}}(\pm,\pm),
\label{cri2}
\eeq
\beq
\du =\Delta\rho^{\mbox{ec}}(+,-) + 
\Delta\rho^{\mbox{be}}_{\mathrm{LUB}}(+,-),
\label{cri1}
\eeq
\beq
\d\rho = \dl - \du = \Delta\rho^{\mbox{be}}_{\mathrm{LUB}}(\pm,\pm) -
\Delta\rho^{\mbox{be}}_{\mathrm{LUB}}(+,-),
\label{difi}
\eeq
where $\Delta\rho^{\mbox{be}}_{\mathrm{LUB}}(\pm,\pm)$ and 
$\Delta\rho^{\mbox{be}}_{\mathrm{LUB}}(+,-)$ are the terms
due to the BE interference simulated
with  LUBOEI. 
As can be seen,  $\d\rho$ resolves the comparatively large difference
between $\Delta\rho^{\mbox{be}}_{\mathrm{LUB}}(\pm,\pm)$ and
$\Delta\rho^{\mbox{be}}_{\mathrm{LUB}}(+,-)$, rather than the
distortion for  like-charged particles alone. The effect depends on the
amount of change in the unlike-charged particle spectra and
other implementations of the  BE interference
may show  a different  effect for $\delta\rho$
than that observed  from the LUBOEI model.
(In the limit that the  BE interference
would not  change the unlike-charged spectra,
the signal would be as much as two to three
times stronger for $\delta\rho$
than that observed  from the LUBOEI model.)
It would be 
interesting to apply other BE simulations based, for example, on 
local \cite{sjo1} and global \cite{be1} re-weighting methods 
or on the LUND string model \cite{be2,be2t}. 

Note that the color reconnection effect  cannot  be detected using 
(\ref{dif}) if it is charge independent, unlike to 
the BE correlations.  However, the color reconnection can be
detected from  unlike-charged  particle combinations, after properly 
removing the correlations from energy-conservation.   

\section{Conclusion} 

One of the main reasons  to study the BE and color reconnection 
effects is the possibility to determine the precision with which 
the W mass can  be measured at LEP2 energies. Moreover, such 
investigations provide an opportunity for probing the structure of 
the QCD vacuum and the details of hadronization.

In this paper we  discuss model-independent signatures of the BE 
and color reconnection effects beyond single-particle spectra.  
The problem of the two-particle correlations in the WW system,  
however, is not as simple as it looks at a first glance:  for WW 
events without 
color reconnection or BE correlations, there are  energy-momentum and 
other correlations which can distort the observed two-particle densities.  
These correlations should be properly taken into account
before estimating the interference effects. We propose to calculate
the difference $\d\rho$, in which contributions from energy-conservation
cancel. 
This differs from the method used  by DELPHI.
Formally, the latter resembles
the traditional way of the BE correlation study, but any quantitative
interpretation of strength and radius parameters and comparison
to values obtained from previous
BE analysis is  misleading.
The method proposed here is expected to be less sensitive to the
distortion from other dynamical correlations between W's. 

As a final remark, we note that  even if the 
energy-conservation effects are properly removed 
as in (\ref{dif}), the absence of a  signal is not a sufficient condition 
for the absence of interference at the hadronization scale. The correlations 
between hadrons originating  from different W's may well exhibit themselves 
in higher-order inclusive  distributions.

\newpage
\section*{Appendix}
\label{glpr}
\appendix 

Recently it was suggested \cite{t2} that the color 
reconnection  effect can lead a smaller mean hadronic  
multiplicity $\la n_{\WW}\ra$ in fully hadronic decay 
than twice the hadronic multiplicity $\la n_{\W}\ra$ 
in semi-leptonic decay, i.e. $\Delta$ in (\ref{aww}) is negative. 
At the present level of statistics at LEP2, no such an effect
has been found \cite{d2,eru}.  A Monte Carlo simulation of BE correlations
based on the Lund  Fragmentation Model gives no support to 
the  possible experimental signal involving single-particle spectra  
as well \cite{be2t,lube}.

Statistically, of course, $(\ref{aww})$ is not
a condition for stochastic dependence between two systems.  
Purely independent production of W bosons has to lead 
to the factorization of the generating 
functionals as in (\ref{den}). This can be illustrated
by replacing auxiliary function $u(p)$  by a constant $z$.
Then the generating functional is reduced to the generating function
$G^{\WW}(z)=\sum (1+z)^nP_n$ for the probabilities
$P_n$ of detecting $n$ hadrons in fully hadronic four-jet WW decays
\beq
G^{\WW}(z)=G^{\Wp}(z)\,  G^{\Wn}(z),  
\label{cal1}
\eeq
\beq
F^{\WW}_q=\frac{\partial^q G^{\WW}(z)}
{\partial z^q}\mid_{z=0},
\label{cal1x}
\eeq
where $G^{\W}(z)$ is the generating function for final-state 
particles in the two-jet WW decays and
$F_q^{\WW}$ is the factorial moment.   

If there is 
a stochastic dependence between W's, (\ref{cal1}) 
has to be modified. However, rigorous information  about
the interdependence  
is necessary to make any definite statement about the exact form of 
$G^{\WW}(z)$. This information is not available because of 
many unknown factors.  One may assume that $G^{\WW}(z)$ can
still be represented by $G^{\Wp}(z)$ and $G^{\Wn}(z)$ if the  
distortions caused by such a dependence  are
not very strong. Then,
\beq
G^{\WW}(z)=G^{\Wp} (z) G^{\Wn} (z) + g(z),
\label{cal2}
\eeq
where $g(z)$ is a function representing possible stochastic
dependence between decay  products  of different W's. 
To preserve the total normalization $G^{\WW}(z=-1)=1$, 
one should put $g(z=-1)=0$, so that $g(z)$ is not a generating 
function in the ``usual'' probabilistic  sense. 
In addition, one must require $g^{'}(z)\mid_{z=0} = 0$ 
and that the form of $g(z)$ cannot
lead to $P_n<0$.
Such a method was used in \cite{chh} to introduce a stochastic
dependence between  Bernoulli sources in order to
modify a positive-binomial distribution.

It is easy to see that $g(z)$ contains  integrated properties of
interference and other effects leading to the dependence of different W's. 
For the average multiplicity $\la n_{\WW}\ra=F^{\WW}_1$ 
in four-jet WW decay, one has from (\ref{cal2})
\beq
\la n_{\WW}\ra =  \la n_{\Wp}\ra +  \la n_{\Wn}\ra 
\label{cal3}
\eeq
since $g^{'}(z)\mid_{z=0}=\Delta=0$. 

For the second-order factorial moment, one obtains
\beq
F_2^{\WW}=F_2^{\Wp} + F_2^{\Wn}
+2\la n_{\Wp}\ra\la n_{\Wn}\ra + g^{''}(z)\mid_{z=0}.
\label{f3}
\eeq
By comparing this expression with (\ref{ff1}), one can see
that 
\beq
\Delta F_2 =g^{''}(z)\mid_{z=0}.
\eeq
If it happens that $g^{''}(z)\mid_{z=0}=0$, then we shall not be  able to 
detect the BE correlations and color reconnection.  
If this is so,   
higher-order factorial moments
(or inclusive densities)   would have
to be checked before one is able to exclude interference effects.

\section*{Acknowledgments}
One of us (S.V.C) acknowledges the hospitality
of  the High Energy Physics
Institute Nijmegen (HEFIN, The Netherlands) 
where this work was started.
We thank B. Buschbeck, T.~Sj\"ostrand, \v{S}. Todorova-Nov\'a and 
A. Tomaradze for helpful discussions.

{}

\newpage

\newpage

\newpage

\end{document}